\UseRawInputEncoding
\documentclass[aps,prl,twocolumn,10pt,amsmath,amssymb,bibnotes,superscriptaddress,longbibliography]{revtex4-1}

\usepackage{graphicx, textcomp}
\usepackage{siunitx}
\usepackage[colorlinks,urlcolor=blue,citecolor=blue,linkcolor=blue]{hyperref}
\newcommand{\g}{\ensuremath{\Gamma}}
\newcommand{\dt}{\ensuremath{\Delta T}}

\newcommand{\di}{\ensuremath{\Delta I_{CS}}}

\newcommand{\iheat}{\ensuremath{I_{bias}}}

\newcommand{\vp}{\ensuremath{V_D}}
\newcommand{\vpc}{\ensuremath{V_P}}
\newcommand{\vc}{\ensuremath{V_T}}

\newcommand{\hotres}{chamber 2}

\newcommand{\zeroone}{\ensuremath{0\to 1}}
\begin{document}

\title{A robust protocol for entropy measurement in mesoscopic circuits}

\author{Tim Child}
\email{timjchild@gmail.com}
	\affiliation{Stewart Blusson Quantum Matter Institute, University of British Columbia, Vancouver, British Columbia, V6T1Z4, Canada}
	\affiliation{Department of Physics and Astronomy, University of British Columbia, Vancouver, British Columbia, V6T1Z1, Canada}
\author{Owen Sheekey}
	\affiliation{Stewart Blusson Quantum Matter Institute, University of British Columbia, Vancouver, British Columbia, V6T1Z4, Canada}
	\affiliation{Department of Physics and Astronomy, University of British Columbia, Vancouver, British Columbia, V6T1Z1, Canada}
\author{Silvia L\"{u}scher}
	\affiliation{Stewart Blusson Quantum Matter Institute, University of British Columbia, Vancouver, British Columbia, V6T1Z4, Canada}
	\affiliation{Department of Physics and Astronomy, University of British Columbia, Vancouver, British Columbia, V6T1Z1, Canada}
\author{Saeed Fallahi}
	\affiliation{Department of Physics and Astronomy, Purdue University, West Lafayette, Indiana, USA}
	\affiliation{Birck Nanotechnology Center, Purdue University, West Lafayette, Indiana, USA}
	\affiliation{Microsoft Quantum Lab Purdue, Purdue University, West Lafayette, Indiana, USA}
\author{Geoffrey C. Gardner}
	\affiliation{Birck Nanotechnology Center, Purdue University, West Lafayette, Indiana, USA}
	\affiliation{Microsoft Quantum Lab Purdue, Purdue University, West Lafayette, Indiana, USA}
    \affiliation{School of Materials Engineering, Purdue University, West Lafayette, Indiana, USA}
\author{Michael Manfra}
	\affiliation{Department of Physics and Astronomy, Purdue University, West Lafayette, Indiana, USA}
	\affiliation{Birck Nanotechnology Center, Purdue University, West Lafayette, Indiana, USA}
	\affiliation{School of Electrical and Computer Engineering,  Purdue University, West Lafayette, Indiana, USA}
    	\affiliation{School of Materials Engineering, Purdue University, West Lafayette, Indiana, USA}
	\affiliation{Microsoft Quantum Lab Purdue, Purdue University, West Lafayette, Indiana, USA}
\author{Joshua Folk}
\email{jfolk@physics.ubc.ca}
	\affiliation{Stewart Blusson Quantum Matter Institute, University of British Columbia, Vancouver, British Columbia, V6T1Z4, Canada}
	\affiliation{Department of Physics and Astronomy, University of British Columbia, Vancouver, British Columbia, V6T1Z1, Canada}
\date{\today}

\begin{abstract}
Previous measurements utilizing Maxwell relations to measure change in entropy, $S$, demonstrated remarkable accuracy of measuring the spin-1/2 entropy of electrons in a weakly coupled quantum dot. However, these previous measurements relied upon prior knowledge of the charge transition lineshape. This had the benefit of making the determination of entropy independent of measurement scaling, but at the cost of limiting the applicability of the approach to relatively simple systems. To measure entropy of more exotic mesoscopic systems, a new analysis technique can be employed; however, doing so requires precise knowledge of the measurement scaling. Here, we give details on the necessary improvements made to the original experimental approach and highlight some of the common challenges (along with strategies to overcome them) that other groups may face when attempting this type of measurement.

\end{abstract}

\maketitle
\section{Introduction}
Direct measurements of entropy in nanoscale systems have the potential to identify and explore exotic quantum states that are otherwise difficult to distinguish from more conventional quantum states. Although entropy, $S$, is a common metric in macroscopic systems obtained through the measurement of heat capacity of the system, this quantity is immeasurably small for nano-scale quantum systems, requiring a different approach entirely.   Strategies have been proposed for quantifying entropy based on electronic measurements of conductance, thermopower, or charge detection, each of which can provide easily detectable signals even in the smallest of quantum devices.\cite{Hartman.2018,Kleeorin.2019,Pyurbeeva.2021x0i, gehring2021, Rozen.2021}  Comparing the three approaches, strategies based on conductance and thermopower can typically be performed closer to equilibrium, but are more limiting in the coupling required between mesoscopic circuit and leads.

Here, we focus on the third approach, using Maxwell relations to measure $S$ (or, more accurately, changes in $S$) by sensing changes in charge with temperature.  Recently, Ref.~\onlinecite{Hartman.2018} followed the charge sensing approach to measure $\Delta S$ associated with the addition of a single spin-1/2 electron in a lithographically-defined quantum dot.    This experiment served as a promising step towards a direct entropy measurement protocol based on charge sensing.  Unfortunately, the precise implementation of the Maxwell relation in that initial work limited its applicability to relatively simple systems for which the measurement of entropy holds little scientific value, and allowed only the determination of entropy change caused by adding one full electron at a time.  At the same time, the experimental method described in that work left room for artifacts in the measurement signal that could contaminate the determination of $\Delta S$.

The goal of this paper is to outline improvements made to the experimental approach in Ref.~\onlinecite{Hartman.2018} that make it more robust at a technical level and applicable to a broader range of measurements.  From the analytical side, the extraction of $\Delta S$ is based on a different formulation of the Maxwell relation \cite{Sela2019},
\begin{equation}\label{eq:eran}
    \Delta S_{\mu_{1}\to \mu_{2}} = \int_{\mu_{1}}^{\mu_{2}}\frac{dN(\mu)}{dT}d\mu,
\end{equation}thereby enabling a determination of entropy (change) continuously as a function of gate voltages or other parameters that control $\mu$. In the experiment, significant improvements to the thermal design and measurement protocol eliminate many sources of error.   Beyond the description of the new experimental protocol, we aim to describe common challenges and possible strategies to overcome them that other groups may encounter in attempting this type of measurement.

\begin{figure}
    \includegraphics[width=1.0\columnwidth]{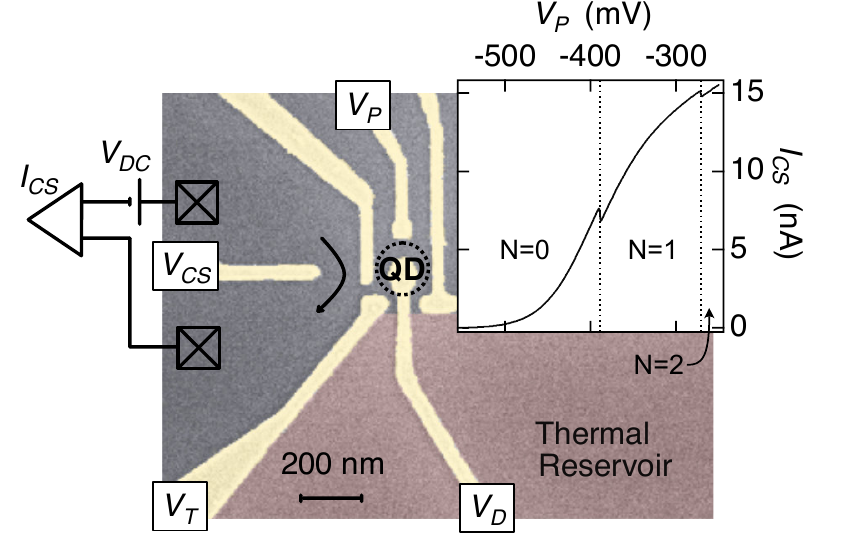}
    \caption{\label{fig:fig1}
    (main panel) False coloured scanning electron micrograph (SEM) of the key parts of the entropy sensor. Electrostatic gates (gold) define the circuit in a 2D  electron  gas  (2DEG).  The thermal electron reservoir (red) can be rapidly heated by driving current through QPCs far away (Fig. \ref{fig:fig3}). (inset) Current through the charge sensor, $I_{CS}$, for a wide sweep of the coarse plunger gate, \vpc, demonstrating the alignment of the \zeroone transition at the steepest part of the trace to maximize sensitivity. 
    }
\end{figure}

\section{Device Design and Layout}

A circuit designed to measure the entropy of a quantum dot (QD) using $\partial S/\partial \mu = \partial N/\partial T$ must have three elements: a QD coupled to an electron reservoir with a tunable chemical potential $\mu$, the ability to change the temperature of this reservoir, and a charge sensor to detect the occupation of the QD.  At the outset, it is important to note that $\mu$ referred to in the Maxwell relation is the chemical potential of the thermodynamic reservoir, which cannot easily be controlled in the experiment\cite{Pyurbeeva.2021x0i}.  Instead, we tune the energy of the QD level, that is, the energy required to add an additional electron to the dot. In practice, it is the difference between $\mu$ and the QD energy that controls when electrons will enter the dot, so tuning the dot level effectively tunes $\mu$ as it appears in Eq.~\ref{eq:eran}.

Figure~\ref{fig:fig1} shows an example of a circuit with the three elements listed above; it is 
similar in functionality to the one described in Ref.~\cite{Hartman.2018}.
The circuit was defined using electrostatic gates on a GaAs/AlGaAs heterostructure, following a standard nanolithography process.
Measurements were carried out in a dilution refrigerator at temperatures ranging from 30 mK up to 500 mK. Electron temperatures below 30 mK were not attainable in our cryostat, and temperatures higher than 500 mK brought in sources of $S$ unrelated to those of interest in the measurement.

The QD itself was defined following standard design guidelines developed through two decades of few-electron dot measurements across the mesoscopics community, see e.g. Refs.~\cite{Elzerman.2004, hanson2007spins,Barthel.2010}.  The gates immediately surrounding the dot were cooled down with a +450 mV bias so that, when cold, the 2DEG under the gates was depleted of carriers with no voltage applied.  The resulting QD could be readily tuned to an occupation of $0 \to \sim 5+$ electrons using the surrounding gates, with \vpc~ dedicated to coarse tuning of the dot occupation.

Two advantages of a charge sensing measurement that simplify device design are that the entropy detection protocol itself is relatively insensitive to coupling through the QD tunnel barrier and that only one such barrier is required.  However, some factors are especially important for the entropy measurement that may not be as relevant in other experiments.  For example, tuning of the chemical potential (QD energy level) is central to this technique, and this tuning must be accomplished without changing other dot parameters significantly.  For this reason, the design includes a gate extending into the middle of the dot, labelled $V_D$ in Fig.~\ref{fig:fig1}, with a very large electrostatic coupling to the QD electron wavefunction: the lever arm of this gate (the ratio of the change in QD energy to the gate voltage applied) was typically 0.2$e$ for this gate, compared with 0.04$e$ or less for $V_P$.


The quantum point contact operated as a charge sensor (CS) was formed by the three leftmost gates in Fig. \ref{fig:fig1}, and used to detect the occupation of the QD \cite{Field.1993, sprinzak2002charge, elzerman2003few}. A DC bias, typically between 50 and 300 \textmu V, was applied across the CS, with the resulting current $I_{CS}$ recorded using a current-voltage converter ($10^8$ A/V, 1 kHz bandwidth set by a two-stage low pass filter).  For the measurement protocol described here, real-time monitoring of the current is important, so the output of the current preamplifier was fed into an analog-digital converter with 2.5 kHz sample rate.  

$V_{CS}$ was tuned to maximize CS sensitivity to charge in the QD.  The inset to Fig.~\ref{fig:fig1} shows 0$\to$1$\to$2 electron transitions for the QD, in this case driven by \vpc, with the 0$\to$1 transition tuned to the steepest slope below the 1st conductance plateau.  The relatively large cross-capacitance between \vpc~and the CS is apparent in the data in Fig.~\ref{fig:fig1} inset: just 200 mV applied to $V_P$ can tune the QPC from pinch-off nearly to full transmission.  This highlights the importance of tuning dot occupation with $V_D$ during the entropy measurement.

The CS sensitivity could also be increased by tuning the gates around the dot to bring the centre of the electron wavefunction as close as possible to the CS: in some cases we were able to achieve a 10\% change in CS transmission due to the addition of an electron to the QD using this gate geometry.  We point out, however, that increasing the QD-CS coupling has both advantages and disadvantages.  Stronger coupling reduces the bias that must be applied to the CS for the same signal-to-noise.  At the same time, stronger coupling shifts the charge detection process farther from the weak-measurement limit that may be desirable from the point of view of back-action on the quantum system under study\cite{Aleiner.1997, silva2003peculiarities, kang2007entanglement}.  Which of these factors is more important will, in general, be different from experiment to experiment.

The coupling of the QD to the heated electron reservoir was controlled by \vc.  Entropy measurements were possible deep into the weakly-coupled regime (very negative \vc), with the limitation that the tunnelling rate, \g, has to be much faster than the measurement rate, that is, the inverse of the time spent sitting at each setting of $\mu$ during which an average $N$  was recorded.  From a thermodynamic perspective, this restriction ensures the QD can transition between all available microstates within the measurement time. In the opposite, strongly-coupled, limit, the primary restriction comes from a strongly reduced $\partial N/\partial T$ signal at high \g.  In our device, $\partial N/\partial T$ approached the noise limit of the measurement when $\g\gtrsim$ 200 \textmu eV $\sim 20k_B T$.


\section{Measurement Protocol}

The measurement of $\partial N/\partial T$ that is central to Eq.~\ref{eq:eran} was carried out by evaluating the discrete derivative $\Delta N/\Delta T$, using the CS to monitor the change in $N$ between two nearby temperatures $\Delta N = (N(T+\Delta T)-N(T))$.
The choice to measure at two particular values of $T$, rather than the simpler approach of oscillating $T$ (approximately) sinusoidally at frequency $f_T$, then locking in to variations in $N$ at $f_T$, was found to be important to the quantitative determination of $\partial T$ in Eq.~\ref{eq:eran} to better than 10-20\% error, and was also helpful in troubleshooting spurious changes in the dot potentials that may appear when attempting to change only $T$.

It is the response of $N$ to temperature alone, with all other parameters (such as $\mu$) constant, that contains information about the entropy of the system.  This requirement, for identical $\mu$ between the two temperatures, turns out to pose a significant experimental challenge.  In practice, any changes $\delta\mu$ in the dot energy between the measurements at $T+\Delta T$ and $T$ will introduce inaccuracy in the entropy measurement by an amount of order $\delta\mu/(k_B \Delta T)$.  For measurements below 100 mK, where $\Delta T$ will then be less than a few 10's of mK, this restricts $\delta\mu$ between the two temperatures to be much less than 1 \textmu eV for an accurate determination of $\Delta S$.  There are both intrinsic and extrinsic factors that must be taken into account in order to keep $\mu$ constant to such a high degree.

\begin{figure}
    \includegraphics[width=1.0\columnwidth]{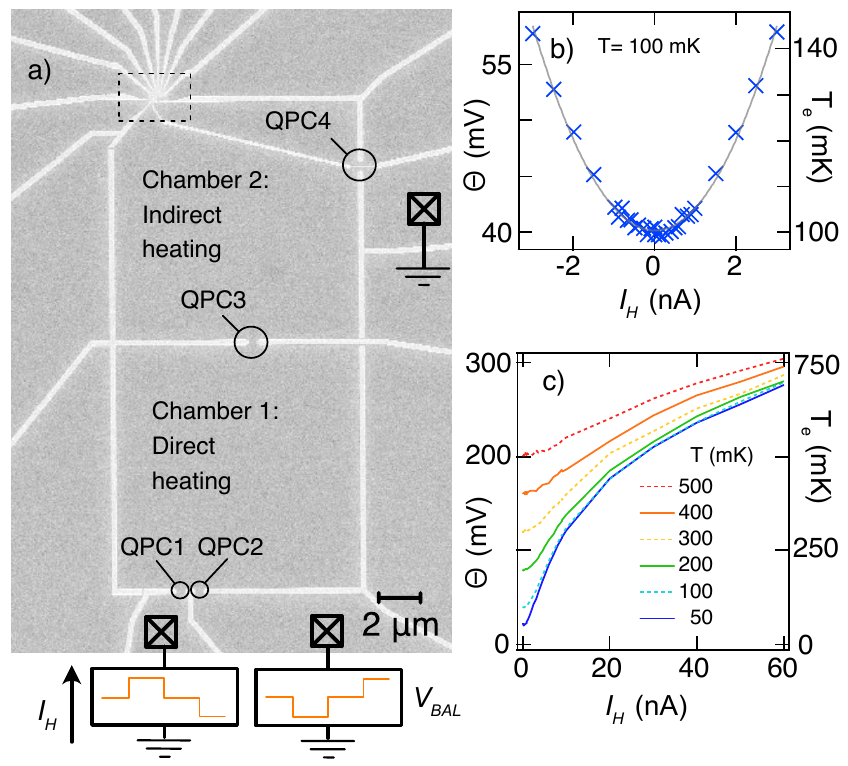}
    \caption{\label{fig:fig2} (a) SEM micrograph of the full measurement device showing the large (10 \textmu m square) chambers used for electron thermalization, QPCs 1 and 2 through which Joule heating current $I_H$ flowed, and QPCs 3 and 4 through which heat diffused but no net current flowed. The dashed rectangle in the upper left is the region shown in Fig.~\ref{fig:fig1}a, including QD and charge sensor.
    (b) Crosses: Broadening of the charge transition ($\Theta$, left axis), converted to electron temperature ($T_e$, right axis),  increases above the sample temperature, $T_s$=100 mK, due to \iheat driven through QPCs 1 and 2. Solid line: quadratic fit to $|I_H|<1$ nA data, with deviations seen at higher $|I_H|$. 
    (c) Extension of panel (b) to higher $I_H$, and for a range of different sample temperatures.  Sub-linear behaviour at very large $I_H$ reflects electron-phonon cooling at higher temperatures.
         }
 \end{figure}

Quantum dots fabricated in GaAs/AlGaAs heterostructures experience intrinsic, albeit small, electrostatic fluctuations due to nearby charge motion in the dopant layer of the heterostructure,  resulting in noise in the QD energy with a frequency spectrum typically between $1/f$ and $1/f^2$\cite{BuizertPRL,liang2020reduction}. It is therefore crucial that the measurements $N(T+\Delta T)$ and $N(T)$ be carried out as close to each other in time as possible, protecting the measurement from noise in the low-$f$ limit. 
The requirement to alternate rapidly between hot and cold reservoir temperatures mandates that the temperature change is accomplished locally on the chip, rather than by changing the temperature $T$ of the entire cryostat.  For this reason, and to minimize the heat capacity of the thermal system, Joule heating due to a bias current $I_H$ was used to raise the electron temperature $T_e$ of the thermal reservoir adjacent to the QD  (Fig.~\ref{fig:fig1}) above the sample (chip) temperature $T$: $T_e=T$ when $I_H=0$  and $T_e=T+\Delta T$ at finite $I_H$.

Driving $I_H$ directly into the thermal reservoir will generally change its potential, however. Since $\mu$ is defined with respect to the chemical potential of the reservoir, this direct effect of $I_H$ must be avoided. At the same time, the advantage of very local heating must be balanced by the requirement for full thermal equilibration of charge carriers in the reservoir, in contrast to the non-equilibrium distribution that is expected when injecting carriers at high bias through a mesoscopic circuit.

A two-chamber heater was used to ensure a thermalized electron reservoir with $\mu$ that did not change when the Joule heating current is applied (Fig.~\ref{fig:fig2}a):  $I_H$ was sourced through QPC1 and drained through QPC2 to heat the first chamber directly, whereas the second chamber (the thermal reservoir immediately adjacent to the QD) was heated indirectly by electrons diffusing from the first chamber through QPC3.  Cooling of the reservoirs was via electron-phonon coupling (especially at higher temperatures) and by diffusion through QPC1, QPC2, and QPC4 to the 2DEG regions connected to ohmic contacts, which remain at the chip temperature due to their large volume and therefore strong electron-phonon equilibration.  For most experiments, QPCs 1, 2, and 3 were set at their 2$e^2/h$ conductance plateaux, while QPC4 was set at 6$e^2/h$.

One advantage of using quasi-enclosed chambers for heating is the relatively low values of $I_H$ required to achieve a significant temperature rise.  Fig.~\ref{fig:fig2}b shows that $T_e$ of \hotres, measured via the broadening of a weakly-coupled charge transition in the QD, can be increased from a sample temperature $T$=100 mK, to $T+\Delta T$=130 mK, with $I_H$ less than 3 nA.  At a quantitative level, of course, the temperature rise for a given current depends on the settings of all 4 QPCs.

The electron temperature is approximately quadratic in $I_H$ for small heating currents, as might be expected from Joule heating power $P\propto I_H^2$, but already by $\Delta T\sim 20$ mK small deviations are visible in Fig.~\ref{fig:fig2}b, where $T$=100 mK.   The deviations become more extreme at higher $I_H$ or lower chip temperature $T$.  Non-quadratic behaviour results from the temperature dependence of the thermal conductivity $\kappa$ between the reservoir electrons and the cold thermal ground, whether it be via electron-phonon coupling to the chip's lattice ($\kappa_{e-ph}\propto T^{3-4}$ expected) or Wiedemann-Franz cooling ($\kappa_{WF}\propto T$ expected) to the cold reservoirs connected to ohmic contacts\cite{mittal1996electron}.   Fig.~\ref{fig:fig2}c illustrates the extreme deviation from quadratic behaviour for large $I_H$, corresponding to large $\Delta T$.  The sub-linear lineshape of $T_e(I_H)$ at the highest currents demonstrates that phonon cooling has become dominant.  It is worth noting that the deviation from $\Delta T\propto I_H^2$ makes the lockin-based approach, which relies on $T_e$ changing at the second harmonic of a sinusoidal $I_H$, especially challenging to calibrate accurately and provides further support for the discrete alternation between $T$ and $T+\Delta T$ used here.

 


 

\begin{figure}
         \includegraphics[width=1.0\columnwidth]{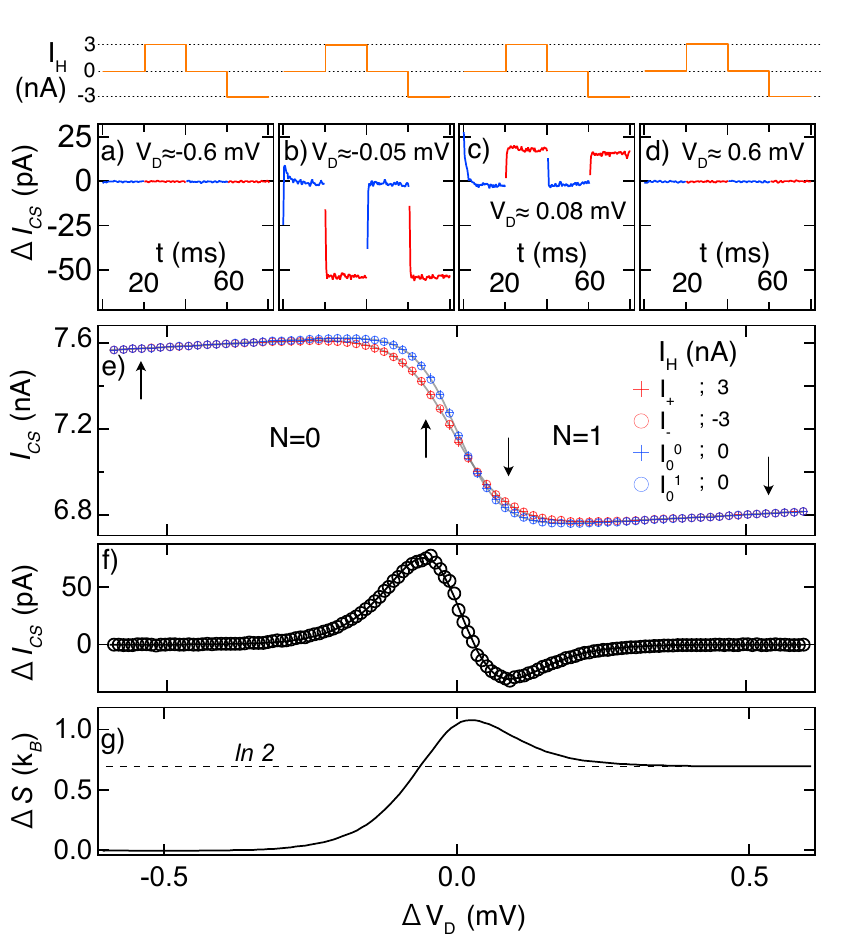}
         \caption{\label{fig:fig3}  
         A step-by-step inspection of the analysis procedure that goes into an eventual calculation of $\Delta S$. The fine-tuning plunger gate, \vp, is used to lower the energy of the QD level such that one electron enters from the thermal reservoir.
         (top) Schematic illustration of $I_H(t)$ through one complete 80 ms cycle. 
         (a-d) Charge sensor current through the 80 ms cycle, calculated with respect to the unheated sections, at four locations on the \zeroone~transition: $V_D=-0.6,-0.05,0.08,0.6$ mV. Data shown here is averaged over 2000 square wave cycles. Blue(red) indicate times at which the thermal reservoir is unheated(heated). The relaxation time of the measurement ($\sim 4$ ms) is visible in panels b,c.
         (e) Charge sensor current separated into averages over the 4 parts of the square heating wave, where heating is applied with an alternating current direction ($I_+$, $I_-$), with zero bias applied in between ($I_0^0$, $I_0^1$). Fits to the average ``cold'' and ``hot'' data are shown in grey. 
         (f) The difference in charge sensor current between the ``cold'' and ``hot'' traces. 
         (g) $\Delta S(\vp)$ obtained by integration of \di~ using Eq. \ref{eq:eran}. $\dt$ is 25.9 mK, equivalent to $0.98$ mV when converted to effective gate voltage, determined from the difference in thermal broadening of heated and unheated $I_{CS}$.
         }
 \end{figure}

The potential of chamber 2 (Fig.~\ref{fig:fig2}a) was held constant by biasing $I_H$ through QPC1 while applying a balancing voltage $V_{BAL}$ behind QPC2  $V_{BAL}$ was tuned such that the potential in \hotres, when sensed directly by the QD, remained constant.  The inverse signs of $I_H$ into QPC1 and $V_{BAL}$ behind QPC2 are illustrated schematically at the bottom of Fig.~\ref{fig:fig2}a.  In order to alternate the temperature while ensuring that $\mu$ stays constant, opposing three-level square waves were created by two channels of a 2.5 kHz digital-analog converter to generate $I_H$ and $V_{BAL}$.  The top row of Fig.~\ref{fig:fig3} shows the square wave driving $I_H$, with an inverse wave setting $V_{BAL}$.   The square wave has four 20ms segments: two segments heated with equal magnitude but opposite sign ($I_H=\pm 3$ nA in this case), separated by two segments at $I_H=0$.  Only after confirming the expected response in all four segments is it possible to conclude that the heating process has not affected $\mu$.


Panels a-d in Fig.~\ref{fig:fig3} show the response of the charge sensor ($I_{CS}$) to the square wave, at the four positions along the charge transition indicated by arrows in Fig.~\ref{fig:fig3}e.  Before and after the transition (Figs.~\ref{fig:fig3}a,\ref{fig:fig3}d) there is no effect of $I_H$.  Checking these ``control" positions is important to confirm the absence of spurious coupling between $I_H$ and the charge sensor, such as capacitive coupling between the wires carrying $I_H$ and those carrying $I_{CS}$, or between the current path of $I_H$ and the charge sensor itself.  Before the midpoint of the transition (Fig.~\ref{fig:fig3}b), Joule heating causes $I_{CS}$ to drop, reflecting extra charge in the dot and therefore positive $dN/dT$ (Eq.~\ref{eq:eran}).  Within the noise of this measurement there is no difference between positive and negative $I_H$.  This confirms that first-order effects of $I_H$ have been strongly suppressed, for example by properly setting $V_{BAL}$.  For $V_D\gtrsim 3$ mV, $dN/dT$ changes sign, before returning to zero well past the transition. 

The raw data, $I_{CS}(t)$, is processed to determine a single value $\Delta I_{CS}$ for each $V_D$. This involves separating the data into two segments corresponding either to $T$ or $T+\Delta T$.  Before that is done, however, it is important to remove the time periods during which the measurement is settling to the new temperature value.  This settling time is clearly visible in Figs.~\ref{fig:fig3}b and \ref{fig:fig3}c, as the $\sim 3$ ms diagonal step preceding each segment at fixed temperature.  We note that the rate of settling is limited by the response of the cryostat wiring in our case; thermal equilibration times within the device are many orders of magnitude faster.  The two segments at $T$, or $T+\Delta T$, are then averaged to find $I_{CS}(T)$, or $I_{CS}(T+\Delta T)$.  These values, determined at each $V_D$, are plotted in Fig.~\ref{fig:fig3}e in blue ($T$) and red ($T+\Delta T$), with the difference, \di, in Fig.~\ref{fig:fig3}f.

$\Delta N/\Delta T$ is obtained from the \di~measurement using parameters obtained from the charge transition itself, $I_{CS}(V_D)$.  Weakly coupled transitions are broadened by the Fermi-Dirac distribution in the reservoir and may be fit to\cite{Houten.1992, Maradan.2014}:
\begin{equation}\label{eq:thermal}
    I_{CS}(V_D) = \frac{-I_e}{2}\tanh\left({\frac{V_D-V_0}{2\Theta}}\right) + I'(V_D-V_0) + I_0,
\end{equation}
where $I_e$ quantifies the sensitivity of the charge sensor to the occupation of the QD, $V_0$ is the centre of the charge transition, $\Theta$ represents the thermal broadening in equivalent gate voltage, $I'$ quantifies the cross-capacitance between $V_D$ and the CS, and $I_0$ is the current through the charge sensor midway through the transition. Although the cross-capacitance is well approximated as a simple linear term for weakly coupled transitions, for more strongly coupled transitions it may have different slopes on the $N=0$ and $N=1$ sides of the transition, which requires more elaborate fitting.

Of these parameters, $I_e$ and $\Theta$ are crucial to the conversion between $\di$ and $\Delta N/\Delta T$.  $I_e$ is the difference in current through the charge sensor between the unoccupied (N=0) and occupied (N=1) states, and is therefore used to scale the charge sensor reading to $\Delta N=-\Delta I_{CS}/I_e$ (the minus sign appears because an increase in N causes a drop in $I_{CS}$).  \dt~is determined from the difference in the broadening term, $\Theta$, for heated and unheated transitions.  This calculation is straightforward when the QD is in the weakly coupled limit, with the charge transition well-modelled by Eq.~\ref{eq:thermal}.   $\Theta$ as determined by fits to Eq.~\ref{eq:thermal} will have units of gate voltage instead of energy, and the lever arm $\alpha\equiv\Delta\epsilon/\Delta V_D$ that converts changes in gate voltage $V_D$ to changes in the dot energy $\epsilon$ would be needed to convert $\Theta$ to $k_B T$.  In practice, it is more convenient to perform the integral in Eq.~\ref{eq:eran} over the gate voltage $V_D$ actually controlled in the measurement, rather than over the equivalent $\mu$ (in units of energy).  Therefore, the denominator in the integrand $\Delta N/\Delta T$ is more conveniently expressed as $\Delta\Theta$ in units of equivalent $V_D$ rather than $\Delta T$ in Kelvin.  The $\Delta S$ obtained by this approach is then in units of $k_B$.
Following this procedure, the factor $\alpha$ cancels and needs not be measured directly.

\section{COMMON PROBLEMS}
\begin{figure}
    \includegraphics[width=1.0\columnwidth]{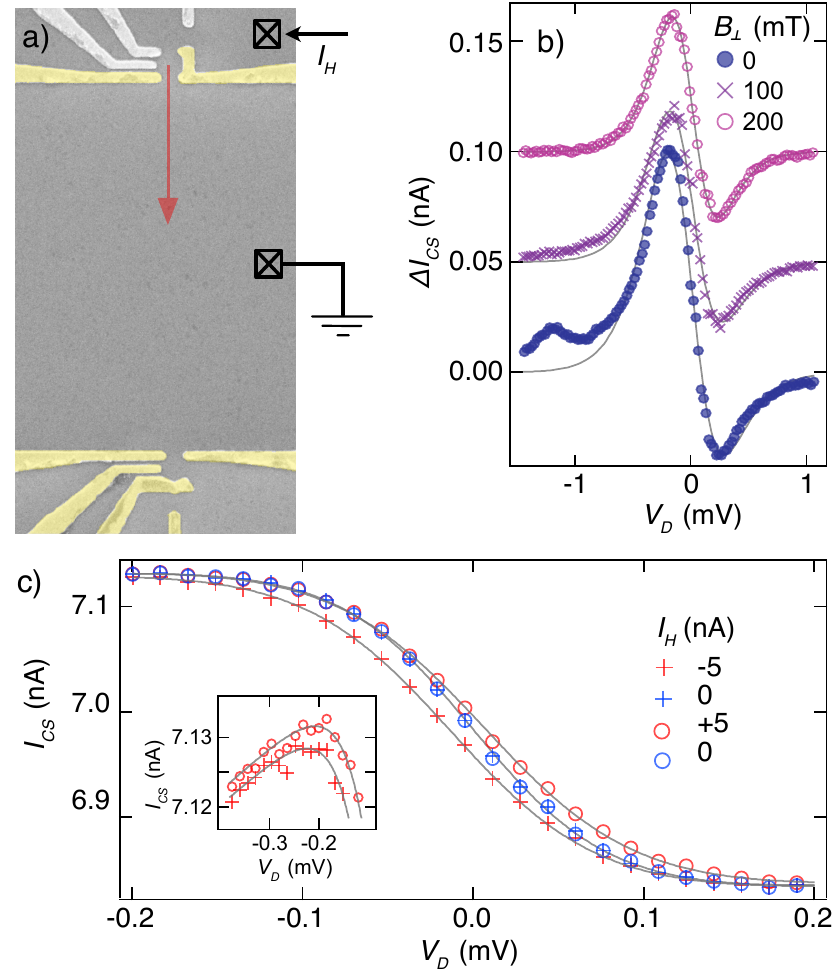}
    \caption{\label{fig:fig4} 
    (a) False-color scanning electron micrograph of entropy measuring circuit from Ref. \cite{Hartman.2018} where the thermal electron reservoir was heated by $I_H$ through a single QPC (top), with no additional confinement of the heated channel.
    (b) Using the circuit in panel a), \di~measurements over the \zeroone~transition for 0, 100, and 200 mT of magnetic field applied perpendicular to the plane of the 2DEG.  100 and 200 mT data are offset by 0.05 and 0.1 nA respectively.   Illustrates the effect of unthermalized electrons from the heater QPC reaching the QD, for 0 and 100 mT data. Fits to theory for weakly-coupled transitions (solid grey) emphasize the deviation of data from theory on the $N=0$ side of the transition.  
    (c) Four segments of $I_H$ square wave averaged separately, analogous to Fig.~\ref{fig:fig3}e and made using the circuit in Fig.~\ref{fig:fig2}a) but without proper balancing to keep the chemical potential of the reservoir at ground.  The result is a shift of $I_H$=+5 nA with respect to -5 nA data. Inset: zoom-in to the $V_D$=-0.4$\rightarrow$-0.1 range of the main panel, showing both lateral and vertical offsets $\pm5$ nA data.
    }
\end{figure}

The rather complicated thermalization device design described in the previous section was arrived at after an initial round of experiments, e.g. Ref.~\onlinecite{Hartman.2018}, with a much simpler design. That design is illustrated in Fig.~\ref{fig:fig4}, with Joule heating through a single QPC directly across a channel from the QD to be measured.  In addition to the more efficient heating in the present design ($\Delta T$=30 mK at $T$=100 mK requires 2.5 nA at 32 \textmu V bias compared to 8 nA at 160 \textmu V bias in Ref.~\onlinecite{Hartman.2018}), Fig.~\ref{fig:fig4}b,c illustrate two of the experimental artifacts that were introduced by the simpler design. 

Figure \ref{fig:fig4}b shows the effect of poor thermalization of the electrons due to $I_H$ before they interact with the dot. Electrons (or holes) passing through the Joule heating QPC enter the reservoir (channel) with very high energy (160 \textmu eV in the example above) compared to the final temperature they will have after equilibration ($k_B\cdot130$ mK$\sim$ 11~\textmu eV).  Due to the ballistic nature of the channel (mean free path $> 5$ \textmu m), they will impinge on the QD far from equilibrium when arriving due to a straight path trajectory\cite{Topinka2001}.  The effect of this non-equilibration is visible in the \di~data taken with transverse field $B_\perp$=0, as a series of bumps preceding the peak in \di, deviating dramatically from the theoretical curve shown with a solid line.  Although we do not have a microscopic explanation for the details of these bumps, they are suppressed by $B_\perp$ as the trajectories from heating QPC are bent away from the QD.  Unfortunately, magnetic fields of at least 200 mT were required to eliminate these deviations entirely (Fig.~\ref{fig:fig4}b), and at this field the entropy measurement was perturbed both by the Zeeman energy of the field and by the onset of Shubnikov de Haas oscillations in the channel. 

Figure \ref{fig:fig4}c illustrates the damaging effect of direct (linear) offset of the reservoir potential due to $I_H$. When $I_H$ is driven through the heater QPC in the geometry from Fig.~\ref{fig:fig4}a, a voltage offset is generated in the reservoir outside the QD due to the non-zero resistance to ground. This offsets $\mu$ in Eq.~\ref{eq:eran}, contravening the requirement to measure $\partial N/\partial T$ with $\mu$ fixed. At the same time, it may have a capacitive effect on the charge sensor, directly affecting the measurement of $N$.  Because these effects reverse with the sign of the current being driven through the heater QPC whereas the Joule heating itself does not, it is easy to identify their influence via a shift of the two heated traces (one at $+I_H$ and one at $-I_H$) away from each other. Direct influence on the reservoir potential causes the traces to separate laterally (Fig. \ref{fig:fig4}c main panel), whereas cross-capacitive effects on the charge sensor cause the traces to separate vertically (Fig.~\ref{fig:fig4}c inset). Averaging the $\pm I_H$ traces together is not sufficient to remove these offsets due to non-linearity in $I_{CS}(V_D)$, and may artificially raise or lower the apparent entropy determined from analyzing \di~data.


\section{OUTLOOK}

We have described an improvement of the thermal circuit design and measurement protocol for quantifying entropy of mesoscopic devices in the quantum limit, based on monitoring how the charge of the system changes with temperature using a Maxwell relation.  To conclude, we offer a few guidelines for how this technique may be improved and made more broadly applicable in the future.

 Extending these measurements to new regimes with higher bandwidth measurements will immediately offer rewards in noise performance. The damaging effect of offset charge noise on quantum dot energy levels, which motivated the fast alternation between temperatures discussed in the Measurement Protocol section, is more severe in this type of experiment than in a typical mesoscopic investigation.  The need to control $\mu$ at a sub-\textmu eV level forces the measurement to be performed at as high a frequency as possible, and fundamental speed limitations involving heat capacity of electrons are orders of magnitude above what was achieved in the present experiment.  

Two more opportunities for improvement stem from the need to remain in thermal equilibrium in order for Maxwell relations to be applicable.  The very act of charge sensing injects a non-equilibrium component into the system dynamics, in principle violating the starting requirement for Maxwell relations. This can be minimized by, first, reducing stray couplings between the sensor circuit and the device under test and, second, by reducing the noise of the charge sensing measurement itself.  At the same time, the theoretical question of how much charge sensing is actually expected to affects $dN/dT$ for a given system remains an important open avenue for study.

At a more mundane level, the requirement for operation in thermal equilibrium is hard to meet in complex circuits, when following the electron heating approach outlined above.  The advantage of heating only electrons is that the heat capacity is minuscule, and temperatures can change rapidly as a result.  The disadvantage is that the electronic system is then out of thermal equilibrium with the phonon lattice, so parts of a multi-component mesoscopic circuit that couple differently to the heated electronic reservoir and to the phonon lattice may end up at different, and poorly defined, effective temperatures.  This concern was not a factor in the proof-of-principle measurement of electron spin entropy laid out in Fig.~\ref{fig:fig1}: the ``system" (the electron in the dot) is easily brought into thermal equilibrium with the reservoir with even the weakest coupling between them because there are no internal degrees of freedom (within $k_B T$) for the first electron in a 200 nm-diameter QD.

For more complex systems, with microstates spaced closely together (but not degenerate) in energy, the challenge will be greater.  Future experiments may ultimately move away from this electron heating approach, to a more intricate thermal circuit that maintains electrons and lattice phonons in thermal equilibrium during the heating step.  This will require a careful design, ensuring that thermal coupling between the chip and cryostat is strong enough to keep the chip close to the base temperature while the heating is off, but weak enough to keep the chip in internal thermal equilibrium during the heating process. 

ACKNOWLEDGEMENTS: This work was undertaken with support from the Stewart Blusson Quantum Matter Institute, the Natural Sciences and Engineering Research Council of Canada, the Canada Foundation for Innovation, the Canadian Institute for Advanced Research, and the Canada First Research Excellence Fund, Quantum Materials and Future Technologies Program.  SF, GCG and MM were supported by the US DOE Office of Basic Energy Sciences, Division of Materials Sciences and Engineering award no. DE-SC0006671, with additional support from Nokia Bell Laboratories for the MBE facility gratefully acknowledged.

\bibliography{draft}

\end{document}